\documentstyle[aps,epsf,multicol,pre]{revtex}
\input epsf
\begin{document}
\draft
\title{Complexity and Fragility in Ecological Networks}

\author{Ricard V. Sol\'e$^{1,2}$ and Jos\'e M. Montoya$^{1,3}$}

\address{$^1$ Complex Systems Research Group, FEN\\
Universitat Polit\`ecnica de Catalunya, Campus Nord B4, 
08034 Barcelona, Spain\\
$^2$Santa Fe Institute, 1399 Hyde Park Road, New Mexico 87501, USA\\
$^3$Department of Ecology, University of Alcal\'a\\
28871 Alcal\'a de Henares, Madrid, Spain}


\maketitle
\begin{abstract}

A detailed analysis of three species-rich ecosystem food webs has shown that 
they display scale-free distributions of connections. Such graphs of interaction 
are in fact shared by a number of biological and technological networks, which 
have been shown to display a very high homeostasis against random removals 
of nodes. Here we analyse the response of these ecological graphs to both random 
and selective perturbations (directed to most connected species). Our results 
suggest that ecological networks are extremely robust against random removal 
but very fragile when selective attacks are used. These observations can have 
important consequences for biodiversity dynamics and conservation issues, 
current estimations of extinction rates and the relevance and definition of 
keystone species.

\end{abstract}


\begin{multicols}{2}

\section{Introduction}

	Ecological research has widely demonstrated that community fragility 
is far from being understood. Issues as which species might be considered as 
specially relevant because of their strong effects on the community have  
lead to a heated debate since Paine's definition of keystone species 
(Jord\'an et al.,1999). Despite this and other discussed topics, it is commonly 
accepted that community fragility is related to how ecological communities 
are structured, specifically to how trophic links are distributed 
throughout the community (May 1974; Pimm 1991). But both the scarcity 
of high-quality data (Polis, 1991; Cohen et al., 1993; Williamson and Martinez, ) 
and the lack of methods 
suitable for a detailed analysis of the complexity of food web organization 
(Cohen et al., 1993) leads to a lack of an unified picture of 
community fragility. 
A number of questions emerge from these studies: How are dynamic and static 
(graph-level) properties related?; How dependent is ecosystem fragility from 
graph architecture?

Recently, there has been an increasing interest in the organization 
of complex networks. These networks go from technological ones 
(Watts and Strogatz, 1998; Albert et al., 1999), 
to neural (Watts and Strogatz 1998; Amaral et al., 2000) or metabolic 
networks (Jeong et al 2000; Wagner and Fell, 2000). All these networks 
can be represented as a graph 
consisting of a set of nodes and the links connecting them. 

Such complex networks share some topological features, as the 
so-called ``small world'' (SW) behavior (Watts and Strogatz 1998; 
Newman, 2000). Some of these webs also exhibit scale-free (SF) 
distributions of links. Specifically, the frequency of nodes with $k$ connections follows 
a power law distribution $P(k) \approx k^{-\gamma}$, where most units are connected 
with few nodes and very few nodes are highly connected.
 
\begin{figure}
\leavevmode
\epsfxsize=8cm
\epsffile{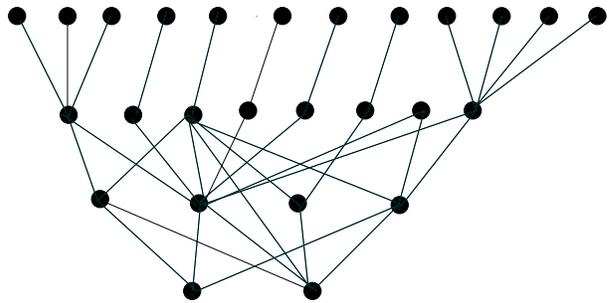}
\vspace{0.5 cm}
\caption{Schematic representation of an ecological graph: few 
species have many links and many have just one or two. This leads 
to fat-tailed (scale free) distributions of connections. This particular 
example follows a power law $P(k)\approx k^{-1.2}$. This picture reflects the 
most common features observed in webs such as those reported here.}
\end{figure}

Networks exhibiting SW properties and SF distributions of connections 
present a characteristic response to the successive removal 
of their nodes, related to the way removals occur (Albert et al 2000; Jeong 
et al 2000). When nodes are removed at random, the network exhibit high homeostasis. 
By contrast,  if most-connected nodes are successively eliminated, the structure 
of the network reveals an intrinsic fragility that eventually leads to a 
breaking into many small subgraphs. This behaviour is not shared 
by other networks, such as purely random ones, where $P(k)$ 
is Poissonian. It has been demonstrated that 
random networks are equally fragile to the way removal of nodes is produced. 

The surprising and general nature of these results immediately 
suggests their application to ecological networks (figure 1), 
which have been recently 
shown to display SF behavior (Montoya and Sol\'e, 2000). Here we examine the 
possible consequences for ecosystem stability against different types 
of species loss. As we will see, these networks display the robustness 
expected for long-tailed distributions of connections but also a high 
fragility against selective species removal.

\section{Food webs analysed}

Because of the limitations of the available data in terms of both 
taxonomic resolution and size (Polis, 1991; Cohen et al., 1993;  
Williams and Martinez, 2000) our study is limited to the three richest 
and best-described food webs available in the ecological literature (figure 2). 
These are: Ythan estuary web (Huxman et al. 1996), 
Silwood Park web (Memmott et al. 2000) and Little Rock Lake web (Martinez 1991). 

\begin{figure}
\leavevmode
\epsfxsize=8cm
\epsffile{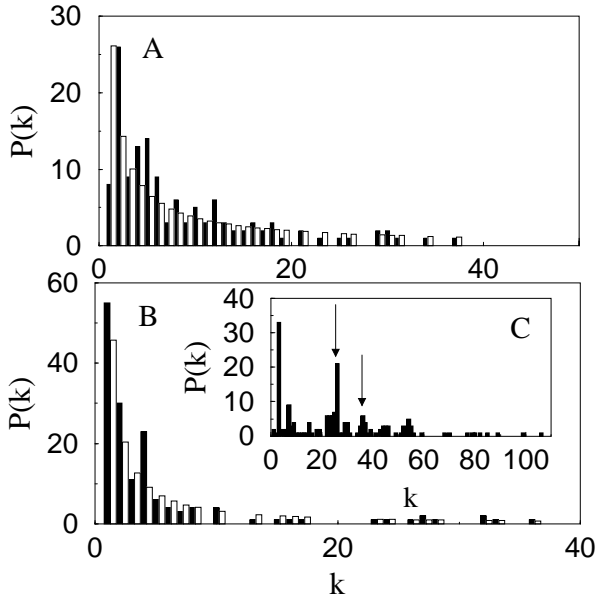}
\vspace{0.5 cm}
\caption{Histograms showing the distribution of links (k) per number of species (P(k)) for the three 
networks analysed here (black bars): (A) Ythan estuary, (B) Silwood web and (C) Little 
Rock lake. Webs A and B are shown together with the best power-law fit (white bars, see table 
I for details). Little Rock web, shown in (C) displays several bumps (two of them 
indicated) due to low-level taxonomic resolution (Montoya and Sol\'e, 2000).}
\end{figure}

    Ythan estuary food web has $N_s=134$ species, being the second largest documented 
web in the UK. Most nodes correspond to real species ($88 \%$) 
while the rest involve 
lower taxonomic resolution (all the species of Acarina or of 
Brown Algae are lumped together in the same node). It is one of the most reviewed 
food webs through ecological literature and the average number of links per 
species is $<k>=8.7$.

    Silwood park food web is a very detailed sub-web (all nodes but one are real 
species) of those species associated with the Scotch broom {\em Cytisus 
scoparius} in a field site of $97$ hectares. The average number of links 
per species is $<k>=4.75$. It includes $154$ species: $60$ predators, $66$ 
parasitoids, $5$ omnivores, 
$19$ herbivores and one plant.

Finally, Little Rock Lake food web corresponds to a small lake. 
It is the largest of the three webs analysed here ($N_s=182$), 
although it has less taxonomic resolution. Only $31\%$ of nodes are 
real species, being most of the nodes genera-level ($63\%$) and the 
rest corresponding to higher taxa. Here $<k>=26.05$.

By using these webs, we have a diverse representation of habitats: 
one food web from a terrestrial habitat (Silwood park), a freshwater habitat (Little 
Rock lake) and an interface environment (Ythan estuary). 
All of them have some features in common. In a previous study, we have shown that 
these networks display small-world properties (Montoya and Sol\'e, 2000). 
In a graph with a small world (SW) topology, nodes are highly 
clustered yet the path length between them is small. In this 
sense a SW stands for a network whose 
topology is placed somewhere between a regular and a totally random distribution of connections. 
These networks display a number of surprising features and have been suggested to 
be of great relevance in different biological contexts (Watts and Strogatz, 1998; 
Jeong et al., 2000; Lago-Fern\'andez et al, 2000).

In general, SW nets have been 
shown to provide fast responses to perturbations and thus provide a 
great source of homeostasis. However, two of these networks also display 
the power law distribution $P(k) \approx k^{-\gamma}$ (figure 2) 
thus belonging to the SF class of networks. Little Rock web is also fat-tailed 
but it displays deviations from the power law due to low taxonomic resolution 
(Montoya and Sol\'e, 2000). For food webs displaying fat-tailed distributions, 
perturbations can have unexpected consequences, which are explored in the 
next section.

\section{Response to species removal}

We have simulated two kinds of species removals: random and directed. 
These correspond to removal of an arbitrary or the highest connected node, 
respectively. Previous studies have shown that the eventual effect of 
removal is network fragmentation, which takes place in very different ways 
depending on the type of removal used (Albert et al., 2000; Jeong et al., 2000). 
Community fragility has been measured in different ways 
in relation to the fraction of species $f$ that have been already removed (Figure 3). 
We have measured the fraction of species contained in the largest species cluster 
$S$ for each $f$; the average size of the rest of the species clusters 
$<s>$ as the food web is being fragmented; and the fraction 
of species becoming isolated due to removals 
of other species on whom their survival depends, which are known in the 
literature as secondary extinctions, 
and can be used as measure of extinction rate (Pimm, 1991). Here $f_c$ indicates 
the fraction of removed species at which the web becomes fragmented into many 
small sub-webs (figures 3a-c).
	
\begin{figure}
\leavevmode
\epsfxsize=8cm
\epsffile{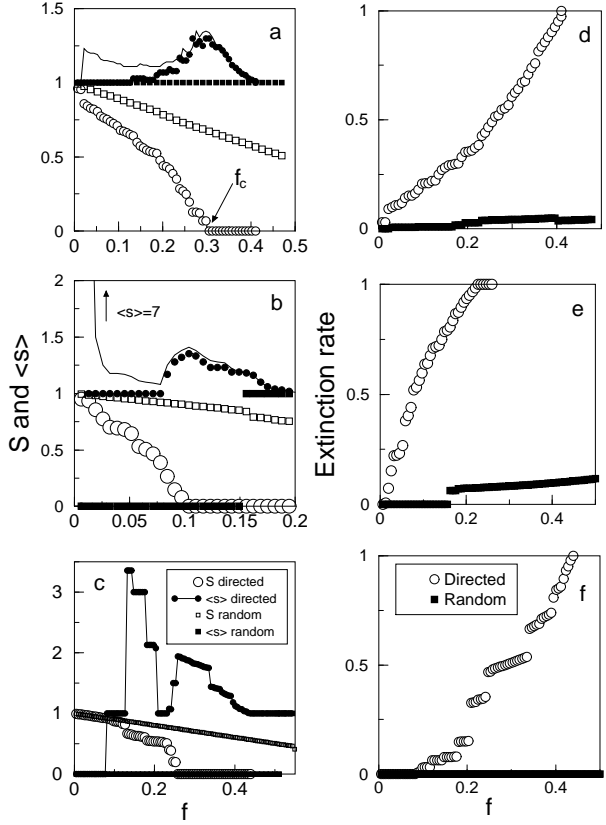}
\vspace{0.5 cm}
\caption{Response of food webs under random (circles) and 
directed (squares) species removals. (a-c): relative size of 
the largest species cluster S (open symbols) and average size of the rest of 
the species clusters $<s>$ (filled 
symbols) in relation to the fraction $f$ of removed species for Ythan estuary, 
Silwood Park web and Little Rock lake food 
web, respectively. Critical thresholds $f_c$ are indicated. In 
(a) and (b), continuous line consider a species cluster that appear 
quickly. This does not rest robustness to the observed trend, 
because it is similar to what happens for $<s>$ and $S$ near $f_c$ if 
that cluster is not considered in the calculations. (d-f): extinction 
rate (fraction of secondary extinctions) 
as a function of $f$ for the webs ordered as before.}
\end{figure}

The nature of the behaviour of the three food webs is very similar 
despite differences in $f_c, S$ and $<s>$ between them. They exhibit high 
homeostasis when random species removals occurs, showing slow, linear decrease in 
the fraction of species contained in the largest cluster $S$. The graph 
cannot be fragmented until extremely high removal has been introduced. This 
can be seen 
in the values of $<s>$, which remain at 0 
(no species clusters different from S), or 1 (due to very few isolated species). But 
what is more revealing is that extinction rates remain at low values even for 
high $f$, so secondary extinctions are almost nonexistent. In fact we can 
estimate the fraction 
of removed species required in order to get food web fragmentation from 
random removal (Cohen et al., 2000): 
$$p_c = 1 - { 1 \over \kappa_0 - 1} \eqno(1)$$
where $\kappa_0$ is estimated from:
$$\kappa_0 = {<k^2> \over <k>} = 
{\sum_{k=1}^K P(k) k^2 \over 
\sum_{k=1}^K P(k) k } \eqno(2)$$
where $K$ indicates the maximum connectivity. This value is shown in table I, 
where we can see that only totally unrealistic removals break the food web. 

\begin{figure}
\leavevmode
\epsfxsize=9cm
\epsffile{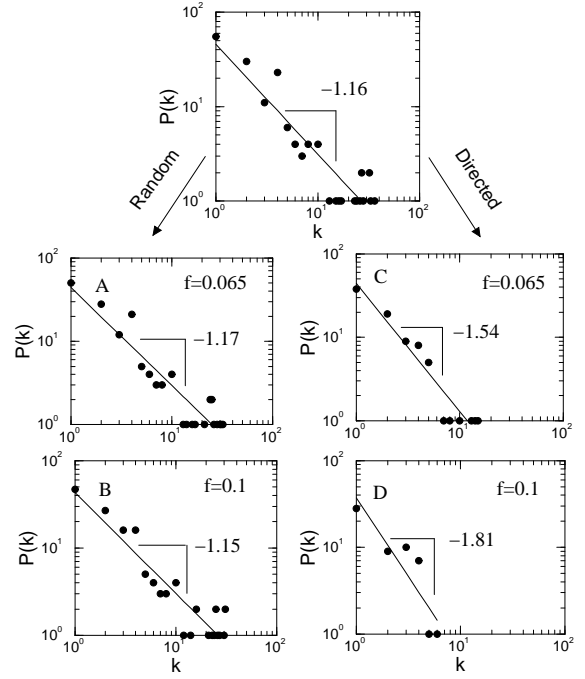}
\vspace{0.5 cm}
\caption{Effects of random and directed removal on the connectivity 
distribution for the Silwood Park web for different fractions of removed 
species: $f=0.065$ (10 removed species), and $f=0.1$ (15 removed species). The 
estimated slopes are indicated ((a) $r^2=0.87; p<0.01$; 
(b) $r^2=0.87; p<0.01$; (c) $r^2=0.92; p<0.01$; (d) $r^2=0.77; p<0.01$.}
\end{figure}

However, what happens when most  
connected species are successively removed is clearly different. These 
webs are extremely vulnerable to such sort of removals. 
This fragility can be seen from: (a) the quick decay of $S$ up to a critical 
fraction of removed species $f_c$ (see table I); (b) the high fragmentation 
of the food webs into species clusters disconnected among them, giving maximum 
local values of $<s>$ at critical points $f_c$ where percolation 
takes place, and (c) the large fraction of 
isolated species occurring at low values of removed species, which reveal 
how fast secondary extinctions will occur (this is specially dramatic 
for our best defined web (Silwood), see figure 3.e). A measure of this phenomenon is 
given by $\pi_c$, the fraction of removed species that leads to an extinction 
rate of one half. Although these are estimations based on a non-directed, 
non-weighted graph, other ingredients will presumably worsen this scenario due to 
other effects derived from indirect interactions (Yodzis, 
1988; Stone and Roberts, 1991; Pimm, 1991) or habitat fragmentation processes 
(Sol\'e and Bascompte, 2001). 

\vspace{0.5 cm}

\begin{center}
\begin{tabular}{||c||c|c|c||} \hline
    & Ythan E. &  Silwood P. & Little Rock \\ \hline 
$<K>$ & 8.71  & 4.78  & 26.15   \\ \hline
$\gamma$ & 1.04  & 1.13  &  - \\ \hline
$p_c$  &  0.94  &  0.93  &  0.97  \\ \hline
$f_c$  &  0.29  &  0.10  &  0.23  \\ \hline
$\pi_c$ & 0.22  &  0.07  & 0.22  \\ \hline
\end{tabular}
\end{center}
\vspace{0.20 cm}
\noindent
\\
Table I: {\em Summary of the average properties of the ecological 
networks analysed (Due to the irregular shape of $P(k)$, 
no good power-law fit for Little Rock Lake can be obtained).
(Here: } (1) $r^2=0.83; p<0.01$; (2) $r^2=0.79; p<0.01$).
\vspace{0.50 cm}

	We can make a simple division of the trophic nature of the 
species for each community into three groups: top predators, 
intermediate species and basal species (Polis, 1991). We find differences 
between analysed food webs in terms of the group that contains more 
highly-connected species, focusing in the set of species that are removed before 
reaching $f_c$. For Ythan estuary, these are mainly 
intermediate species (fishes and invertebrate organisms, $60\%$), few top 
predators (birds, $20\%$) and parasites ($15\%$) that cannot be easily included 
in any of these three groups. For Silwood park, most of these species 
are herbivores ($66\%$), that could be considered as basal species since 
only one plant ({\em Cytisus scoparius}), is present. Hemipterous  
omnivores are also important ($26\%$), but, as happens with parasites at Ythan, 
they do not belong to any of the three trophic groups. Finally, for Little 
Rock no basal nodes are highly connected, 
being intermediate species belonging to zooplankton, benthic 
invertebrates ($70\%$) and top predators (such as fishes, $24\%$) the most 
connected ones.  

	Another fundamental question related to the fragility of these 
communities is how species removals affect the distribution of links per 
species under each kind of simulated attack. In other words, do food webs 
maintain their SF distributions as species are successively deleted? Our 
analysis shows that indeed, SF distributions are stable up to high values of $f$ 
(more than $50 \%$ of species eliminated in the three food webs) when  random 
removals occur, while those particular distributions quickly disappear under 
directed removal. In Figure 4 we have represented this 
for the Silwood web. Under random removal, the long-tailed distributions show 
little variation. In contrast, that SF topology is lost 
when just few highly connected species are removed, which is likely to 
promote ecosystem collapse.

\section{Discussion}

The trophic organization of species-rich communities is similar to other 
complex network topologies (Albert et al 2000, Jeong et al 2000). They are 
extremely heterogeneous, being their topology dominated by few highly 
connected nodes around which the rest of the network is organized, with 
a scale-free distribution of connections. This complex organization entails 
some keys for ecological fragility. We have shown that SF food webs are very 
robust under random deletion of species. Secondary extinctions remain at low 
values because the probability of being removed decreases with $k$ according to
$P(k) \approx k^{-\gamma}$, so it is unlikely that a highly connected species will be 
deleted. Such robustness becomes weakness under removals directed to species 
with many connections. We find specially significant the differences in 
extinction rates between random and directed attacks, raising up to $95$ 
times higher for removals of highly connected (keystone) species.

	All the definitions of keystone species found in ecological 
literature share one feature: keystone species have large effects on other 
species in the community. The set of effects involved in each definition 
are very different. Those effects have been studied mainly qualitatively
 by the removal or introduction of species (see Jordan et al 1999 for a review).
 Quantitative approximations have used simulated food webs (Pimm 1991;  
Jordan et al 1999). In this respect, our approximation to the fragility 
of real, species-rich food webs through topological changes may help to design new 
quantitative methods for a priori identification of keystone species. We can 
identify keystone species as highly connected because of the 
effects of their removal in terms of secondary extinctions. The quick arrival 
to $f_c$ and the high extinction rates corresponding to low values of $f$ when 
those keystone species are removed stresses the importance of identifying 
and protecting highly connected species that maintain the stability of ecological 
communities. 

By using this method, it is the topology of the food web instead of the 
trophic position of species what determines which species are keystone. 
In this respect, not only top predators must be considered as 
keystone species but also other organisms from different trophic levels, 
in agreement with previous studies (Bond 1993; Davic 2000). Making a 
simple division of species into the three classic trophic categories, we have 
seen that keystone species belong to different categories in each of 
the analysed networks. A common feature found in the two  best taxonomically 
described food webs is that species that feed on more than one 
trophic level (omnivorous species and parasites, which in most cases 
could be considered as a special type of omnivory, Polis and Strong 1996) are a representative 
group in the set of keystone species, enhancing the importance of 
omnivorous species in the stability of ecological communities 
(McCann and Hastings, 1997).

Our results suggest that there are basic principles of ecological 
organization (not 
revealed by previous analyses) underlying the assembly process of diverse 
communities. These principles are in fact present at other scales: when a 
given spatial habitat is fragmented, there is a critical percolation 
threshold which leads to the breakdown of the habitat available into 
many small patches (Bascompte and Sol\'e, 1996; Hanski, 1999). This threshold 
has very important consequences for metapopulation persistence and allows to 
define appropriate criteria for conservation in fragmented landscapes 
(Keitt et al., 1997). 

Small-world and SF properties allows high food web complexity in terms 
of biodiversity. By one side, recent studies (Solow and Beet 1998; 
Montoya and Sol\'e 2000) have shown that real food webs are more clustered 
from what would be expected from random wiring. 
These evidences support the hypothesis that compartmentalization (a 
characteristic of SW) is a way to enhance species 
coexistence in species-rich communities (May 1974). Besides, 
species that interact with a great number of species do so weekly. 
Conversely, species with strong interactions interact with few species 
of the community (May 1974, Paine 1992). Data on interaction strengths 
in natural food webs show that these networks are characterized by many 
weak interactions and very few strong interactions (Paine 1992; Raffaeli and 
Hall 1996). Recent studies also suggest that the balance
 of nature is related with those widespread weak interactions 
(McCann et al., 1998; Polis 1998). Thus highly 
connected species are important in promoting community stability and persistence. 
The detailed knowledge of this particular organization could lead to 
an improvement in the understanding of ecological assembly rules 
(Drake, 1990a; Drake 1990b). 

This approximation to community fragility has obvious 
caveats derived from a lack of dynamics and taxonomic limitations. 
Besides, our findings are based on a non-directed, 
and non-weighted graph. Top-down and bottom-up effects due to species 
removals are considered together. A separate knowledge of these effects 
has been previously reported, arguing that whole community effects were 
more interesting and more relevant in the determination of keystone 
species (Jordan et al 1999). Preliminary 
analysis of model ecosystems suggest that our results are robust (Sol\'e and 
Montoya, unpublished). 

Current estimations of extinction rates are based on species-area 
relations, combined with estimates of habitat loss (May et al., 1995). 
The addition of secondary extinctions due to removal of keystone species, 
together with other indirect effects are likely to increase such projections. 
Food webs are described at a local scale, but the estimated extinction 
rates obtained from our study could be important in  
forecasting extinction rates at regional and global scales.

\vspace{0.6 cm}

{\Large Acknowledgments}

\vspace{0.2 cm}

We thank our friends and colleages Javier Gamarra, David Alonso, Ramon Ferrer and 
Miguel Angel Rodriguez for their help at different stages of this work. 
JMM wants to dedicate this work to the memory of his grandfather 
Evaristo Teran for whom nature was living. 
This work has been supported by a grant
CICYT PB97-0693 and The Santa Fe Institute (RVS).

\vspace{0.5 cm}

\begin{enumerate}

\item
Albert, R., Jeong, H. and Barab\'asi, A-L. (2000) Error and attack tolerance of 
complex networks. Nature 406, 378-382.

\item
Amaral, L.A.N., Scala, A., Barthelemy, M. and Stanley, H.E. (2000) Classes of 
small-world networks. Proc. Nat. Acad. Sci. USA, 97, 11149-11152.

\item
Barab\'asi, L-A. and Albert, R. (1999) Emergence of 
scaling in random networks. Science 286, 509-512.

\item
Bascompte, J. and Sol\'e, R. V. (1996) 
Habitat Fragmentation and Extinction Thresholds in spatialy explicit models,  
J. Anim. Ecol. 65, 465-470

\item
Bollob\'as, B. (1985) {\em Random Graphs}. (Academic Press, London) 

\item
Bond, W.J. (1993) Keystone species. Pages 237-253 in E-D. Shultze and H.A. 
Mooney, editors. Biodiversity and ecosystem function. Springer-Verlag, 
Berlin, Germany.

\item
Cohen, J.E. et al. (1993) Improving food webs. Ecology 74, 252-258.

\item
Cohen, R., Erez, K., ben-Avraham, D. and Havlin, S. (2000) Resilience of the 
Internet to random breakdowns. Phys. Rev. Lett. (in press) 

\item
Davic, R.D. (2000) Ecological dominants vs. keystone species: a call for 
reason. Conservation Ecology 4(1): r2 [online] URL: 
http://www.consecol.org/vol4/iss1/resp2

\item
Drake, J. A. (1990a) The mechanics of community assembly rules. 
{\em J. Theor. Biol.}, {\bf 147}, 213-233.

\item
Drake, J. A. (1990b) Communities as assembled structures: do rules
govern pattern?. {\em Trends Ecol. Evol.}, {\bf 5}, 159-163.

\item
Hanski, I. (1999) {\em Metapopulation Ecology}, Oxford U. Press.

\item
Huxman, M., Beaney, S. and Raffaelli, D. (1996) Do parasites reduce 
the chances of triangulation in a real food web?. Oikos, 76, 284-300.

\item
Jeong, H., Tombor, B., Albert, R., Oltvai, Z.N. and Barabasi, A-L. (2000) The 
large-scale organization of metabolic networks. Nature 407, 651-654.

\item
J\'ordan, F., Tak\'acks-S\'anta, A. and Moln\'ar, I. (1999) A reliability 
theoretical 
quest for keystones. Oikos 86, 453-462.

\item
Keitt, T.H., D.L. Urban, and B.T. Milne. (1997)  
Detecting critical scales in fragmented landscapes. 
Conservation Ecology {\bf 1}, 4.

\item
Lago-Fern\'andez, L. F., Huerta, R., Corbacho, F. and Sig\"uenza, J. A. (2000) 
Fast response and temporal coherent oscillations in small-world networks. Phys. Rev. Lett. 
84, 2758-2761. 

\item
McCann, K. and Hastings, A. (1997) Re-evaluating the 
omnivory-stability relationship in food webs. Proc. 
R. Soc. Lond. B. 264, 1249-1254.

\item
McCann, K., Hastings, A. and Huxel, G.R. (1998) Weak 
trophic interactions and the balance of nature. Nature 395, 794-798.

\item
Martinez, N.D. (1991) Artifacts or attributes? Effects of resolution on the 
Little Rock Lake food web. Ecol. Monog., 61, 367-392.

\item 
May, R.M. (1974) {\it Stability and complexity in model ecosystems}. 
Princeton U. Press.

\item
May, R.M., Lawton, J.H. and Stork, N.E. (1995). 
Assesing extinction rates, in {\em Extinction Rates} (Lawton, J.H. 
and May, R.M., eds.) pp. 1-24. Oxford U. Press.

\item
Memmot, J., Martinez, N.D. and Cohen, J.E. (2000) Predators, parasitoids and pathogens: 
species richness, trophic generality and body sizes in a natural food web. 
J. Anim. Ecol., 69, 1-15.

\item
Montoya, J. M. and Sol\'e, R. V. (2000) Small world patters in fodd webs. Submitted 
to J. Theor. Biol. Also: Santa Fe Institute Working Paper 00-10-059. 

\item
Newman, M. E. J. (2000) Models of small worlds: a review. J. Stat. Phys. 
101, 819-841.

\item
Paine, R.T. (1966) Food web complexity and species diversity. American 
Naturalist 100, 65-75.

\item
Paine, R.T. (1992) Food-web analysis through field measurement of per capita 
interaction strength. Nature 355, 73-75.

\item
Pimm, S. L., Lawton, J. H. and Cohen, J. E. (1991) Food web patterns and their 
consequences. {\em Nature (Lond.)} {\bf 350}, 669-674. 

\item
Pimm, S. L. (1991) {\it The Balance of Nature}. Chicago Press.

\item
Polis, G.A. (1991) Complex trophic interactions in deserts: 
an empirical critique of food web theory. American Naturalist, 138, 123-155.

\item
Polis, G.A. and Strong, D.R. (1996) Food web complexity and community dynamics. 
American Naturalist 147, 813-846.

\item
Polis, G.A. (1998) Stability is woven by complex webs. Nature 395, 744-745.

\item
Raffaelli, D.G. and Hall, S.J. in Food Webs: Integration 
of Patterns and Dynamics (eds. Polis, G.A. and Winemiller, K.O.), 
185-191. Chapman and Hall, New York.

\item
Sol\'e, R. V. and Bascompte, J. (2001) {\em Complexity and Self-organization 
in Evolutionary Ecology}, Monographs in Population Biology.
(Princeton U. Press) (to appear)

\item
Stone, L. and Roberts, A. (1991) Conditions for a species to gain 
advantage from the presence of competitors. Ecology 72, 1964-1972.

\item
Yodzis, P. (1988) The indeterminacy of ecological interactions as perceived 
through perturbation experiments. Ecology 69, 508-515. 

\item 
Wagner, A. and Fell, D. (2000) The small world inside large metabolic 
networks, {\em Santa Fe Institute Working Paper} 00-07-041.

\item
Watts, D.J. and Strogatz, S.H. (1998) Collective dynamics of 
"small-world" networks. 
Nature 393, 440-442.

\item
Williams, R. J. and Martinez, N. D. (2000) Simple rules yield complex 
food webs. Nature (Lond.) 404, 180-183

\end{enumerate}

\end{multicols}

\end{document}